\documentclass[seceq]{ptptex}

\usepackage{graphicx}

\title{
Supersymmetric Free-Damped Oscillators: Adaptive Observer Estimation
of the Riccati Parameter }

\author{       
V. \textsc{Ibarra-Junquera} (1), H. C. \textsc{Rosu} (2), O. \textsc{Cornejo-P\'erez} (2)%
}

\inst{     
(1) Faculty of Chemical Sciences, Univ. of Colima, Coquimatl\'an, Col., Mexico,\\
(2) Potosinian Institute of Science and Technology,\\
Apdo Postal 3-74 Tangamanga, 78231 San Luis Potos\'{\i}, Mexico
}

\abst{         
\noindent A supersymmetric class of free damped oscillators with three
parameters has been obtained in 1998 by Rosu and Reyes through the
factorization of the Newton equation. The supplementary parameter is
the integration constant of the general Riccati solution. The
estimation of the latter parameter is performed here by employing
the recent adaptive observer scheme of Besan\c con {\it et al.}, but
applied in a nonstandard form in which a time-varying quantity
containing the unknown Riccati parameter is estimated first. Results
of computer simulations are presented to illustrate the good
feasibility of this approach for a case in which the estimation is
not easily accomplished by other
means.\\ 



\begin{center} {\tt Int. J. Theor. Phys. Vol. 45, No. 11, Nov. 2006, 2109-2117.\\
Rec. Dec. 28, 2005; accept. Feb. 14, 2006; online Sept. 22, 2006.
physics/0411019v2, papptp2.tex} \end{center}}

\begin{document}

\maketitle

\section{Three-parameter Free-damped Modes}

In 1998, Rosu and Reyes \cite{rr} introduced classical {\em
three-parameter} free damped oscillators by means of the
nonrelativistic supersymmetry (factorization) technique.
They showed that in this approach the usual two-parameter free damped oscillator equation (the dots are for time derivatives)
$$
y^{''}+2\beta y^{'} + \omega _{0}^{2}y=0
$$
is generalized to a damped oscillator equation of the form
\begin{equation}  \label{1}
\tilde{y}^{''}+2\beta \tilde{y}^{'} + \omega ^{2}(t; \gamma)
\tilde{y}=0~,\,\, {\rm where}\,\, \quad \omega ^{2}(t;
\gamma)=\Big[\omega _{0}^{2}-\frac{2\gamma ^2}{(\gamma t
+1)^2}\Big]~,
\end{equation}
possessing
damping solutions with singularities that we call here $\gamma$ modes, see below. 
The new damping solutions depend  
on an additional parameter $\gamma$ given by the general Riccati solution associated to the original two-parameter damped equation. This $\gamma$ parameter introduces
a new time scale in the damping phenomena that is related to the time at which the singularities occur.

For the three types of
free damping, the $\gamma$ modes are the following:

(i) Underdamping: $\beta ^{2}<\omega _{0}^{2}$,
$\omega _{u}=\sqrt{\omega _{0}^{2}-\beta ^2}$
\begin{equation} \label{under}
\tilde{y} _{u}= -\tilde{A} _{u}e^{-\beta t}
\Big[\omega _{u}\sin(\omega _{u}t+\phi)+\frac{\gamma}{\gamma t+1}
\cos(\omega _{u}t+\phi)\Big]~.
\end{equation}

(ii) Overdamping, $\beta ^2>\omega _{0}^{2}$,
$\omega _{o}=\sqrt{\beta ^2-\omega _{0}^{2}}$,
\begin{equation} \label{over}
\tilde{y} _{0}=-\tilde{A} _{o}e^{-\beta t}\Big[\omega _{o}
\sinh(\omega _{o} t+\phi)-
\frac{\gamma}{\gamma t +1}\cosh (\omega _{o} t+\phi)\Big]~.
\end{equation}

(iii) Critical damping, $\beta ^2=\omega _{0}^{2}$.
\begin{equation} \label{critical}
\tilde{y}_{c}=e^{-\beta t}\Big[\frac{-A_{c}\gamma}{\gamma t+1}+\frac{D_{c}}{\gamma ^2}
(\gamma t +1)^2\Big]~.
\end{equation}


Damped motion of this type could be quite common in nature, although
if the value of $\gamma$ is very small it is difficult to
distinguish it from the usual two-parameter free damping. In such
cases it is still possible to estimate $\gamma$ through some
powerful estimation methods. The problem of parameter estimation in
dynamical systems has motivated extensive studies of the adaptive
observer designs for linear and nonlinear systems during the last
decade. Among the single output adaptive schemes that can be
directly applied to our case we cite those of
Marino,\cite{Marino90} Marino and Tomei,\cite{Marino95} and Besan\c con.\cite{Besancon04} It
is the purpose of this paper to show that very good estimates for
$\gamma$ can be obtained by applying the latter scheme.
The main reason to use the Besan\c con scheme is its simplicity with
respect to the scheme of Marino that requires solving more
differential equations in real time in order to obtain the
corresponding velocity to recover the unknown value of the parameter(s). In the next section, Eq.~(\ref{1}) is written as a system of two first-order differential equations in the affine form which is required in order to apply Besan\c con's adaptive observer scheme. This is detailed in Section 3 where we also give some pedestrian definitions of the technical concepts involved in the adaptive estimation algorithms. A small conclusion section ends up the paper.

\section{Equivalent System of First Order Differential Equations}

Denoting $X_1=\tilde{y}$ and $X_2 =\dot{\tilde{y}}$, Eq.~(\ref{1})
can be rewritten as
\begin{eqnarray}
\dot{X_1} &=& X_2 \label{eq1} \\
\dot{X_2} &=& -2 \beta X_2 + \left( \frac{2 \gamma^2}{(\gamma t
+1)^2} - \omega _0^2\right) X_1 \label{eq2}~.
\end{eqnarray}

\noindent
We will consider $X_1$ as the naturally measured state (the most
easy to measure). Therefore, it seems logical to take $X_1$ as the
output of the system, that is $y = CX$, where $C=[1\ 0]$ and $X$ is
state vector. In addition, we consider that $\beta$ is available in
some way and only $\gamma$ stands as an unknown parameter. Thus, the
complete monitoring of the system means the on-line estimation of
$X_2$ and the on-line identification of the $\gamma$ value. The
available methods for parameter identification need that the unknown
parameters are in affine form, i.e.
 \begin{eqnarray}
\dot{X_1} &=& X_2 \label{eqna1} \\
\dot{X_2} &=& -2 \beta X_2 + \alpha X_1 \label{eqna2}~,
\end{eqnarray}
where $\alpha$ is the unknown constant parameter.
It is clear that
$\gamma$ is not an affine parameter but we can try to estimate the time-dependent quantity
\begin{equation}\label{alpha1}
 \alpha (t)=\left( \frac{2 \gamma^2}{(\gamma t
+1)^2} - \omega _0^2\right)~,
\end{equation}
because it appears in the affine form. Notice that, even though
$\alpha$ is not a constant,  its evolution is significant only close
to the origin, that is, as time increases $\alpha$  turns rapidly
into a constant parameter.

\section{Design of the Adaptive Observer}

It is well known that only in very special occasions one can have a
sensor on every state variable, and some form of reconstruction from
the available measurement output data is needed, in many cases. In
general, an observer is an algorithm that reconstructs the
unmeasurable states of a system from the measurable output. The
so-called high gain techniques proved to be very efficient for state
estimation, leading to the well-known concept of \emph{high gain
observer} \cite{J.P.Gauthier92}. The gain is the amount of increase
in error in the observer's structure. This amount is directly
related to the velocity with which the observer recovers the unknown
signal. The high-gain observer is an algorithm in which the amount
of increase in error is constant and usually of high values in order
to achieve a fast recover of the unmeasurable states.
However, when a system depends on some unknown parameters, the
design of the observer has to be modified in such a way that the
state variables and parameters could be estimated. This leads to the
so called
\emph{adaptive observers}, i.e., observers that can change in order to work better or
provide more fit for a particular purpose.

Here, we present an interesting application of the recent scheme of
Besan\c con and collaborators,\cite{Besancon04} for unknown
parameters that appear linearly in the dynamical system to the
interesting case
of state and parameter estimation of the three-parameter free-damped
oscillator. In this particular case, the unknown parameter appears
in a nonlinear manner and in order to avoid this difficulty we group
together a set of parameters which leads to a time-dependent
parameter. The block scheme of the adaptive observer for this case
is presented in Fig.~(1).

The main assumptions on the considered class of systems are 
that if all parameters were known, some high-gain observer could be
designed in a classical way, and that the systems are ``sufficiently
excited" in a sense which is close to the usually required
assumption on adaptive systems (the signals must be rich
enough so that the unknown parameters can indeed be identified).

We write the $\gamma$ free
damping motion system given by Eqs.~(\ref{eqna1})-(\ref{eqna2}) in Besan\c con's form:
\newenvironment{mycase}{\left \{\def\arraystretch{1.2} \array{@{}l@{\quad}l@{}}}{ \endarray \ \right. }\makeatother
\begin{equation}
   \text{$\ \ \ \digamma$ =}  \begin{mycase}
\dot{X} = AX + \varphi(X) + \Psi(X) {\theta} \nonumber\\
y = C X  \nonumber
 \end{mycase} 
\nonumber
\end{equation}

\noindent where:

\noindent $A$ - Brunovsky-type matrix, i.e., of the form $A=\delta
_{i,j+1}$, where  $\delta _{ij}$ is the Kronecker symbol

\noindent
$X^n \in
\mathbb{R}$ - $n$-component real vector

\noindent
$y$ - measured output

\noindent $\Psi(t) \in \mathbb{R}^{n \times p}$ - the vector of
known functions ($p$ = the number of unknown parameters)

\noindent
$\theta \in \mathbb{R}^p$ - vector of unknown parameters (i.e.,
a vector belonging to some known compact set
that should be estimated through the measurements of the output $y$).

\medskip

\noindent
For the $\gamma$ system, we have: $n=2$, $p=1$ and $\theta=\alpha$.

\noindent
Therefore the entries of the matrix $A$ are
$a_{11}=0$, $a_{12}=1$, $a_{21}=0$ and $a_{2,2}=0$, the entries
of the vector $\Psi$ are $\psi_{11}=0$ and $\psi_{21}= X_1$,
whereas the entries of the vector $\varphi$ are $\varphi_{11}=0$
and $\varphi_{21}= -2 \beta X_2$. \\

\begin{figure}[x]    
\centering
\includegraphics[scale=0.25]{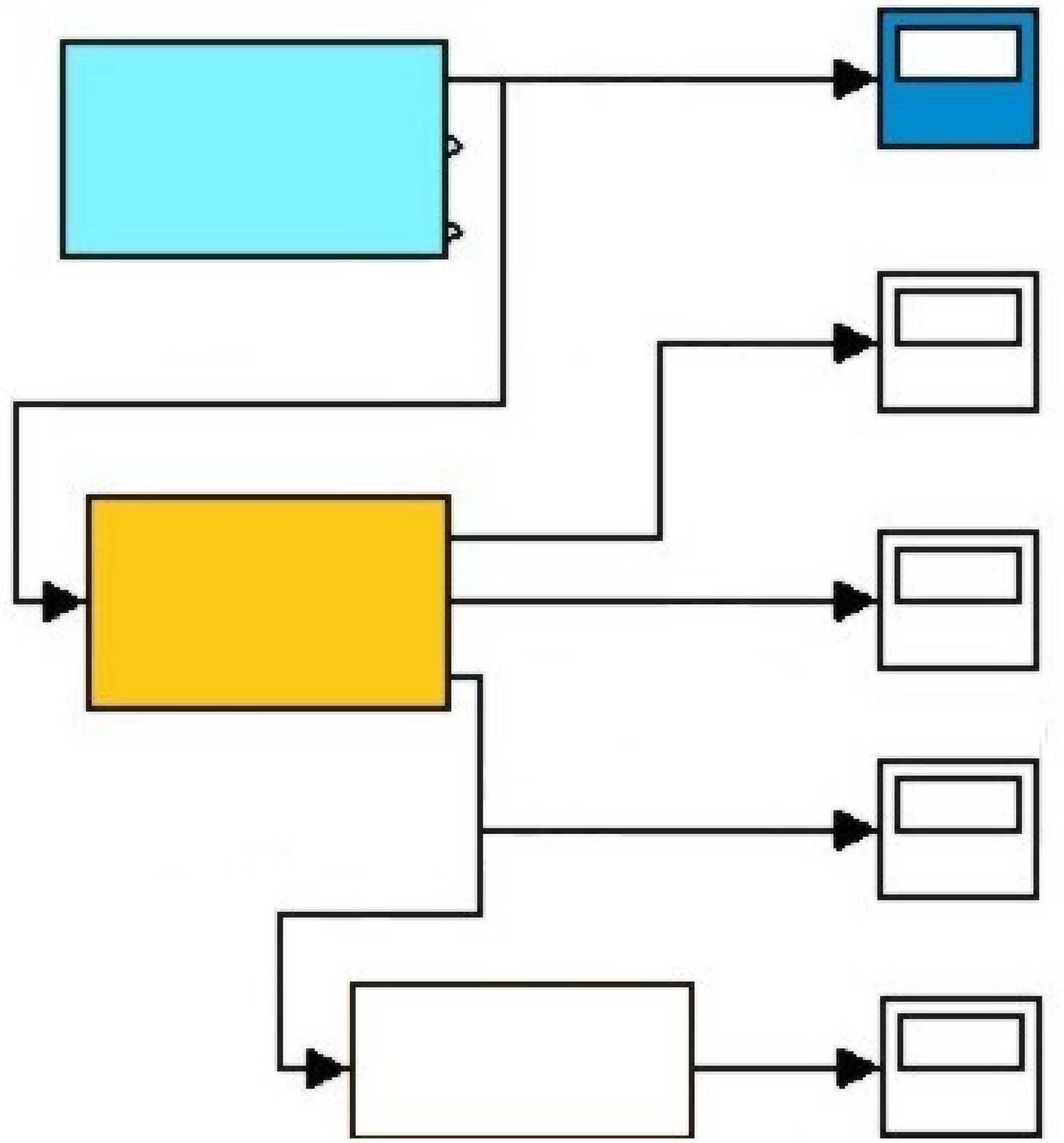}
    \put(-23,115){\small{$X_1$}}
    \put(-39,82){\small{estimated $X_1$}}
    \put(-39,50){\small{estimated $X_2$}}
    \put(-39,22){\small{estimated $\alpha$}}
    \put(-39,-8){\small{estimated $\gamma$}}
    \put(-90,10){\tiny{ rebuilt$\gamma$}}
    \put(-90,132){\tiny{${X}_3$}}
    \put(-90,65){\tiny{$\hat{X}_2$}}
    \put(-90,73){\tiny{$\hat{X}_1$}}
    \put(-149,100){\small{Original System}}
    \put(-123,43){\small{Adaptive}}
    \put(-123,33){\small{Observer}}
\caption{Schematic representation of the adaptive observer, where
the output of the system is the input of the adaptive observer and
the outputs of the latter are the rebuilt states and the parameter
$\alpha$ from which it is possible to rebuild $\gamma$.} \label{Fig2d}
\end{figure}

We consider now the following dynamical system

\begin{equation}
  \text{$\ \ \ \hat{\digamma}$=}  \begin{mycase}
    \dot{\hat{X}} = A \hat{X}+ \varphi(\sigma(\hat{X})) + \Psi(\sigma({\hat{X}})) \sigma({\theta}) + \Lambda^{-1} \left [ \rho K + \Gamma \Gamma^T C^T  \right]  \left(y  - C \hat{X}  \right) \\
    \dot{\hat{\theta}} = \left [\rho^n \Gamma^T C^T \right] \left(y  - C \hat{X}  \right) \\
    \dot{\hat{\Gamma}} = \rho \left( A-K C \right) \Gamma +\rho
    \Psi(\sigma(\hat{X}))~,
    \end{mycase} 
   \nonumber
\end{equation}
\\
\noindent
where 
$\Gamma \in \mathbb{R}^{n \times p}$, $K$ is some vector for which
$A-KC$ is a stable matrix, $\Lambda = {\rm diag}[1, \rho^{-1},
\ldots, \rho^{-(n-1)}]$, $\rho \in \mathbb{R}_{+}$ being a constant
to be chosen, and $\sigma$ is a so-called {\em saturation function}.
The saturation function is a map  whose image is bounded by the
chosen upper and lower limits, $B$ and $b$, respectively
\cite{kh92}. This avoids the over and/or under estimation, keeping
the possible values within the set of the physically reliable
values, and consequently increasing the chance of the quick
convergence to the true values.
Thus, the latter is the following map
\begin{equation}\label{sigma}
\sigma(s)=\begin{mycase}
    B $\quad$ s>B\\
    s $\quad$ b\leq s \leq B\\
    b $\quad$ s<b~.
    \end{mycase} 
   \nonumber
   \end{equation}

 $\hat{\digamma}$ has been proven by Besan\c con to be a
{\em global exponential adaptive observer} for the system
$\digamma$. That means that for any initial conditions $X(t_0)$,
$\hat{X}(t_0)$, $\hat{\theta}(t_0)$ and $\forall \theta \in
\mathbb{R}^p$, the errors $\hat{X}(t)-X(t)$ and
$\hat{\theta}(t)-\theta(t)$ tend to zero exponentially fast when $t
\rightarrow \infty$. Consequently, for the free damping motion
system, the matrix $A- KC$ has the eigenvalues
\begin{equation}
\lambda_{1,2} = -1/2\,k_{{1}}\pm1/2\,\sqrt {{k_{{1}}}^{2}-4\,k_{{2}}}
\end{equation}

\noindent Selecting $k_{2}=\left(\frac{k_{1}}{2}\right)^{2}$, we get
the eigenvalues $\lambda_1=\lambda_2= -k_{1}/2$, and
choosing any $k_1 >0$ we make $A- KC$ a stable matrix. Thus,
the explicit non matrix form of the observer given by $\hat{\digamma}$
is the following

\begin{eqnarray}
\dot{\hat{X}}_1 &=& \hat{X}_2 + \left( \rho k_1 + {\Gamma_1}^2
\right) (X_1-\hat{X}_1) \nonumber\\
\dot{\hat{X}}_2 &=& -2 \beta
\sigma(\hat{X}_2)+\sigma(\hat{X}_1)\sigma(\hat{\theta})+ \rho
\left( \frac{\rho {K_1}^2}{4}+ \Gamma_1 \Gamma_2 \right)(X_1-\hat{X}_1) \nonumber\\
\dot{\hat {\theta} } &=& \rho^2 \Gamma_1 (X_1-\hat{X}_1) \nonumber\\
\dot{\Gamma}_1 &=& \rho (-k_1 \Gamma_1 + \Gamma_2) \nonumber\\
\dot{\Gamma}_2 &=& - \frac{1}{4} \rho {k_1}^2 \Gamma_1 + \rho
\sigma(\hat{X}_1))~.\nonumber
\end{eqnarray}

\noindent
To recover now the value of the parameter $\gamma$ from the
estimated value of $\alpha$, we solve for $\gamma$ the equation $2 [\gamma /(\gamma t
+1)]^2 - \omega _0 ^2=\alpha$, giving

\begin{eqnarray}
\gamma = {\frac {-t(\alpha\,t+{\beta}^{2})-\sqrt
{2\,\alpha+2\,{\beta}^{2}}} {t^2(\alpha + {\beta}^{2})-2}}~.
\end{eqnarray}

\noindent We still can hope that the value of $\gamma$ can be estimated
with sufficiently good accuracy until the blow-up time $t_d=\sqrt{2}/\left( \alpha +
\beta^2\right)^{\frac{1}{2}}$ and indeed this is what we have found for the $\gamma$ modes.


In the computer simulations that we performed for illustrating
purpose we took always $\rho =100$ for the constant parameter
defining the diagonal matrix $\Lambda$.
Figures~(2) - (5)  
show the results of numerical simulations for an underdamped case
with the following parameters: $\beta =0.04$, $\gamma =0.15$,
$k=1.00$, $\omega _0 =1.00$. In figure (6), we present numerical
simulations for the Riccati parameter $\gamma$ for the overdamped
case with $\beta =1.0$ and $\omega _0 =0.2$ in the plot (a) and the
critical case with $\beta =\omega _0=0.1$ in the plot (b). Figure 5a
shows the initial stages of the process of convergence displaying
big oscillations. After this, one can always notice the convergence
to the true value once the estimate enters the neighborhood of the
true value determined by the exponential criterium of convergence to
the neighborhood, where small oscillations around the true value are
quite natural.
The  numerical simulations have been performed with the relative
values of the parameters chosen to illustrate the three classes of
damping, but otherwise there are no restrictions on their values.

\begin{figure}[x]    
  \hfill
  \begin{minipage}[t]{.45\textwidth}
    \begin{center}
      \includegraphics[height=4.4cm]{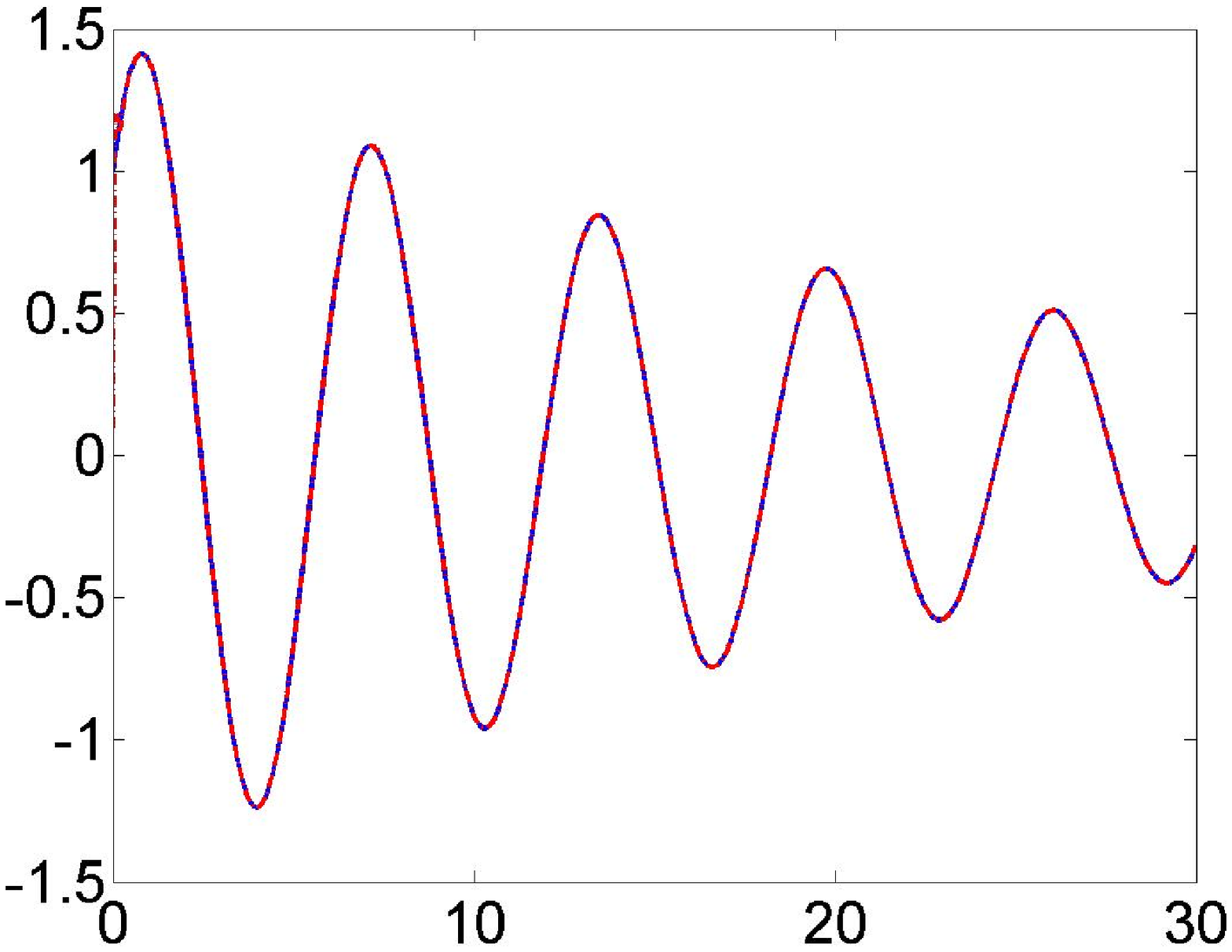}
      \put(-37,-10){time}
       \put(-37,106){(a)}
      \put(-175,10){\rotatebox{90}{$X_1$ }}
    \end{center}
  \end{minipage}
  \hfill
  \begin{minipage}[t]{.45\textwidth}
    \begin{center}
      \includegraphics[height=4.4cm]{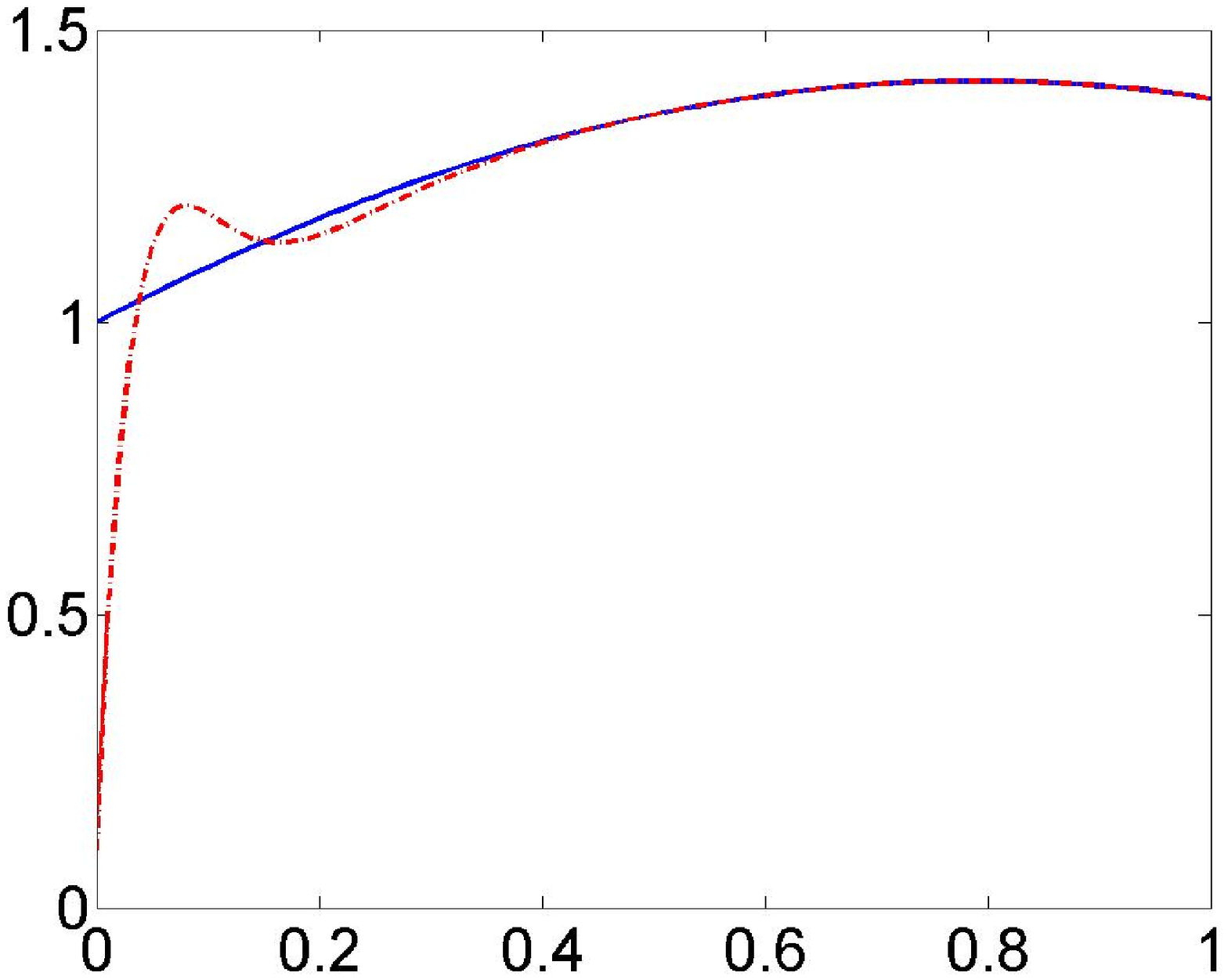}
       \put(-35,-10){time}
       \put(-174,10){\rotatebox{90}{$X_1$ }}
       \put(-37,108){(b)}
    \end{center}
  \end{minipage}
  \hfill
  \caption{Numerical simulations for the state $X_1$ in the underdamping case. In (a) the solid line represents the time evolution of the true state and the dotted
  line gives the evolution of its estimate. 
The plot (b) is a detail of the figure (a) to appreciate the variation  of $\hat{X_1}$ in the beginning.}\label{simulaciones X1}
\end{figure}

\begin{figure}[x]   
  \hfill
  \begin{minipage}[t]{.45\textwidth}
    \begin{center}
      \includegraphics[height=4.4cm]{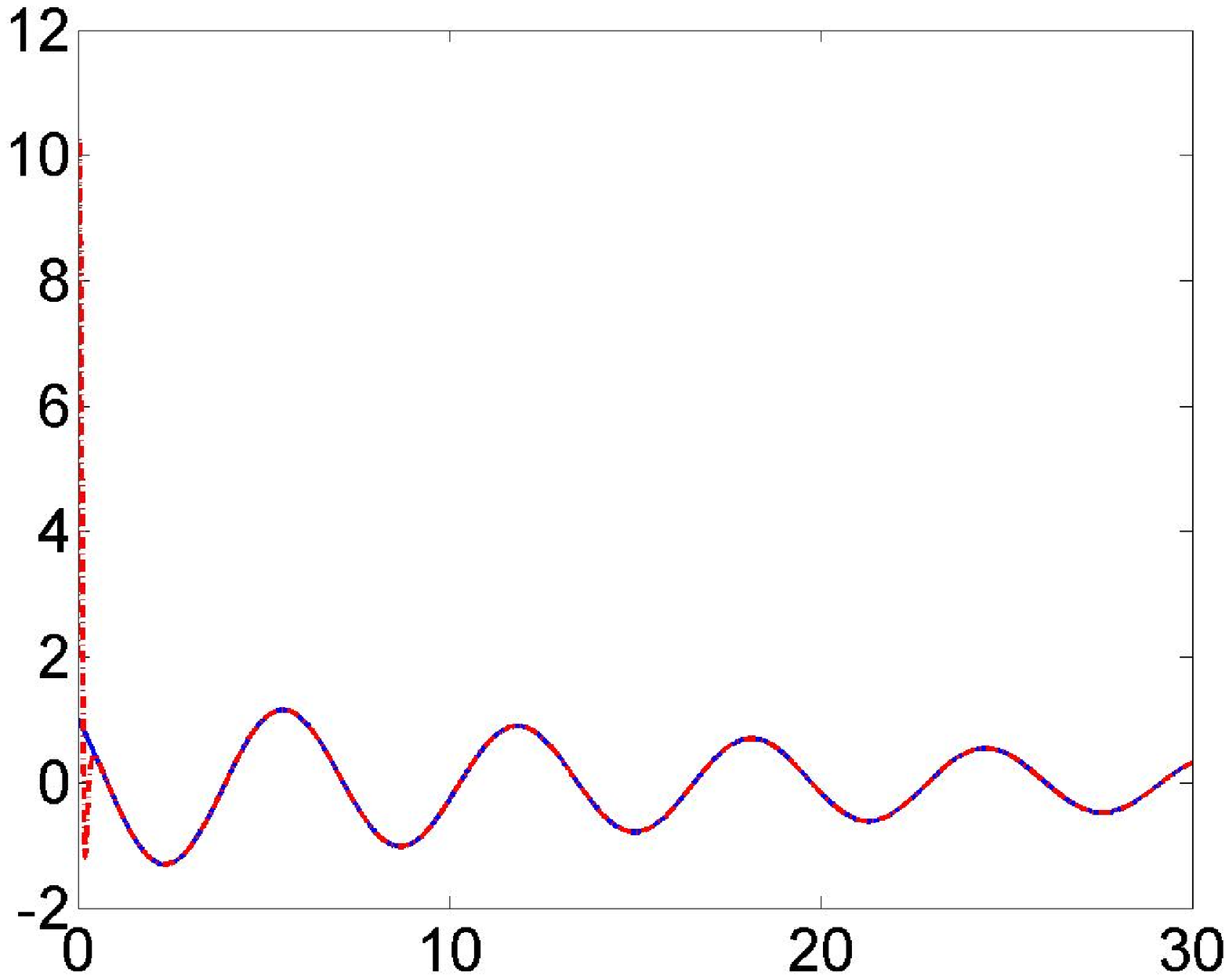}
      \put(-37,-10){time}
       \put(-32,106){(a)}
      \put(-175,10){\rotatebox{90}{$X_2$ }}
    \end{center}
  \end{minipage}
  \hfill
  \begin{minipage}[t]{.45\textwidth}
    \begin{center}
      \includegraphics[height=4.4cm]{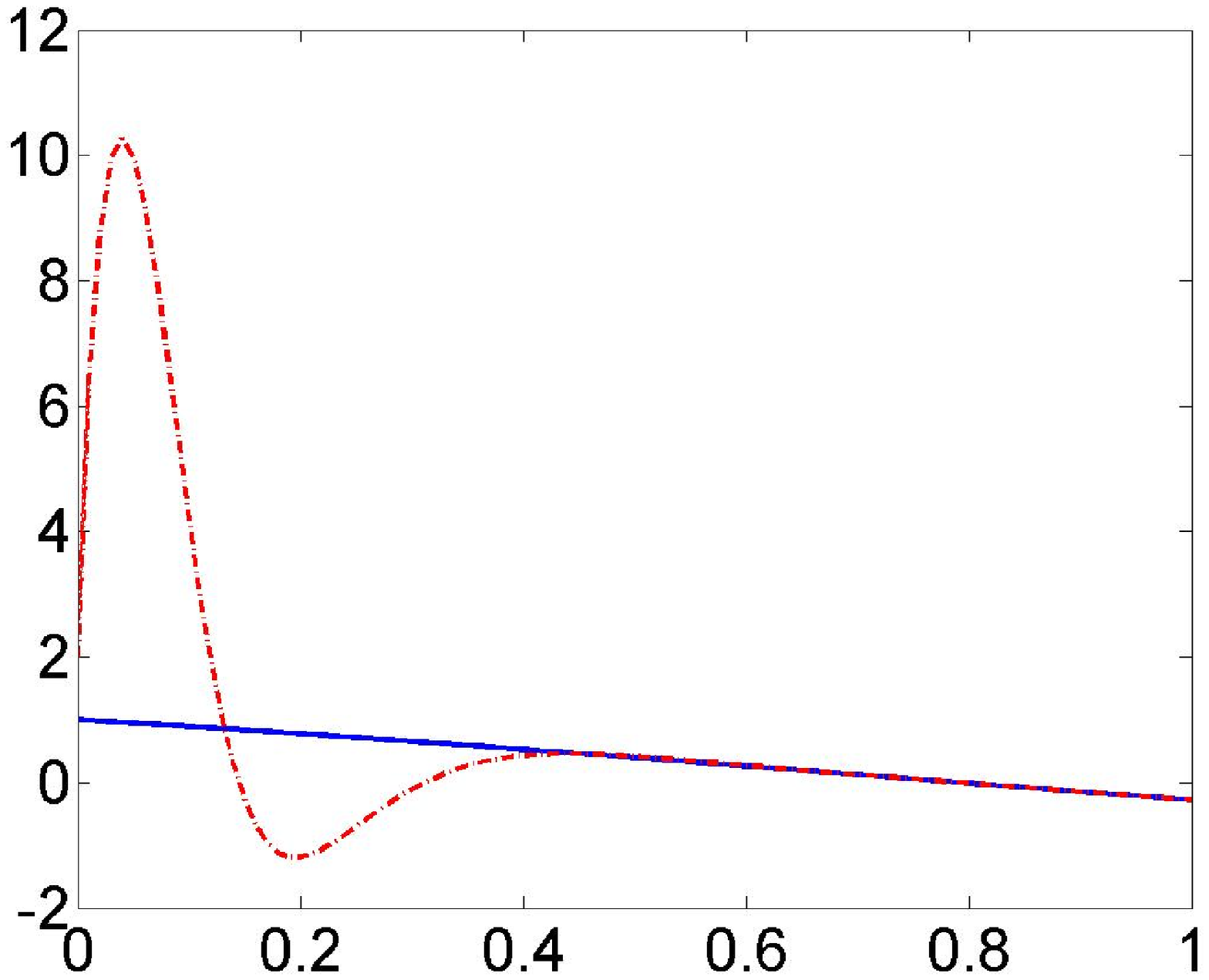}
       \put(-174,10){\rotatebox{90}{$X_2$ }}
         \put(-37,-10){time}
       \put(-32,106){(b)}
    \end{center}
  \end{minipage}
  \hfill
  \caption{Numerical simulation for the state $X_2$ in the underdamping case. In (a)
the solid line represents the time evolution of the true state and the dotted
  line gives the evolution of its estimate. 
Plot (b) is a detail of (a) to see better the variation  of
   $\hat{X_2}$ in the beginning.}\label{simulaciones X2}
\end{figure}

\begin{figure}[x]  
  \hfill
  \begin{minipage}[t]{.45\textwidth}
    \begin{center}
      \includegraphics[height=4.2cm]{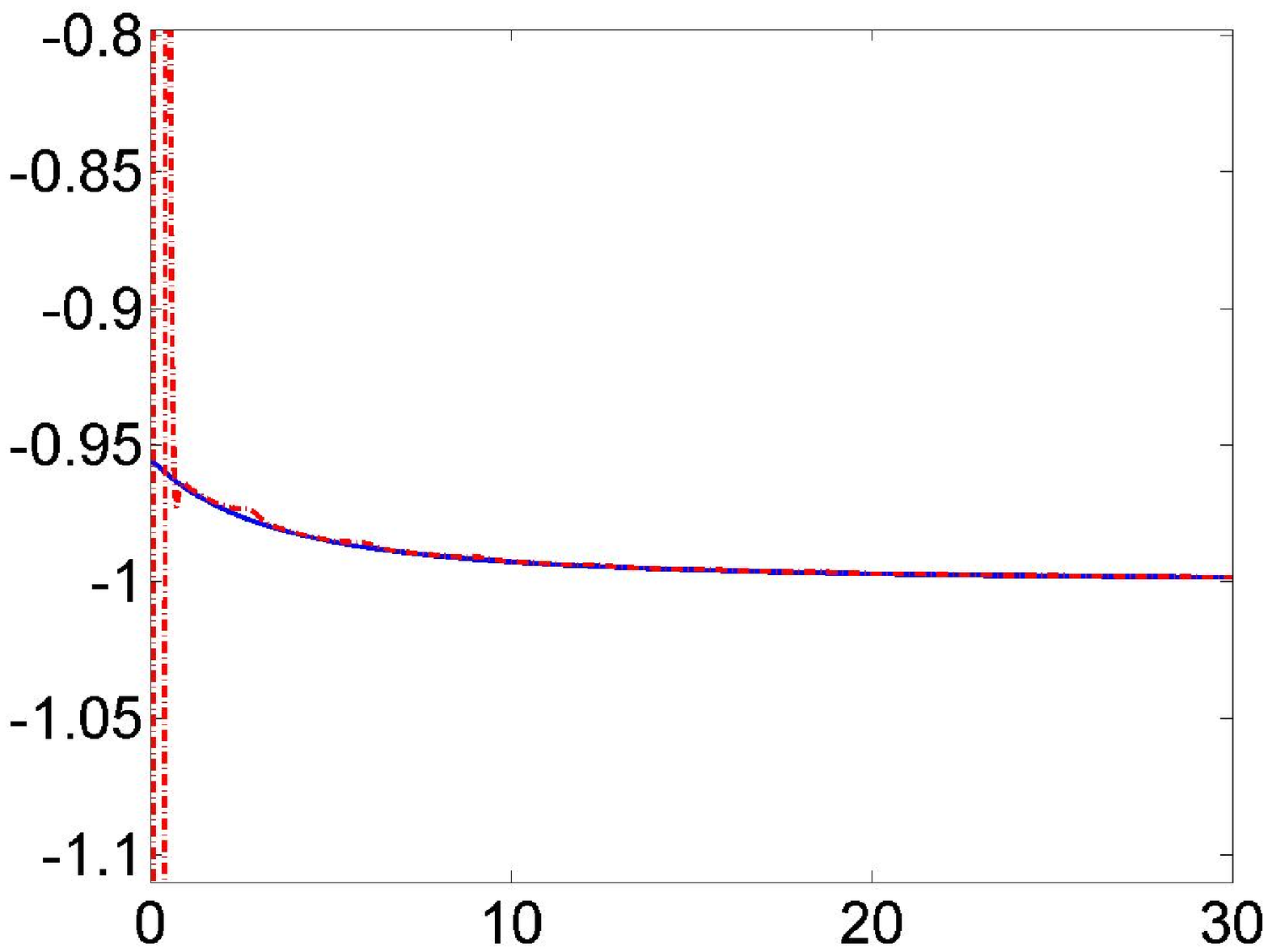}
      \put(-37,-10){time}
       \put(-38,90){(a)}
      \put(-165,10){\rotatebox{90}{parameter $\alpha$ }}
    \end{center}
  \end{minipage}
  \hfill
  \begin{minipage}[t]{.45\textwidth}
    \begin{center}
      \includegraphics[height=4cm]{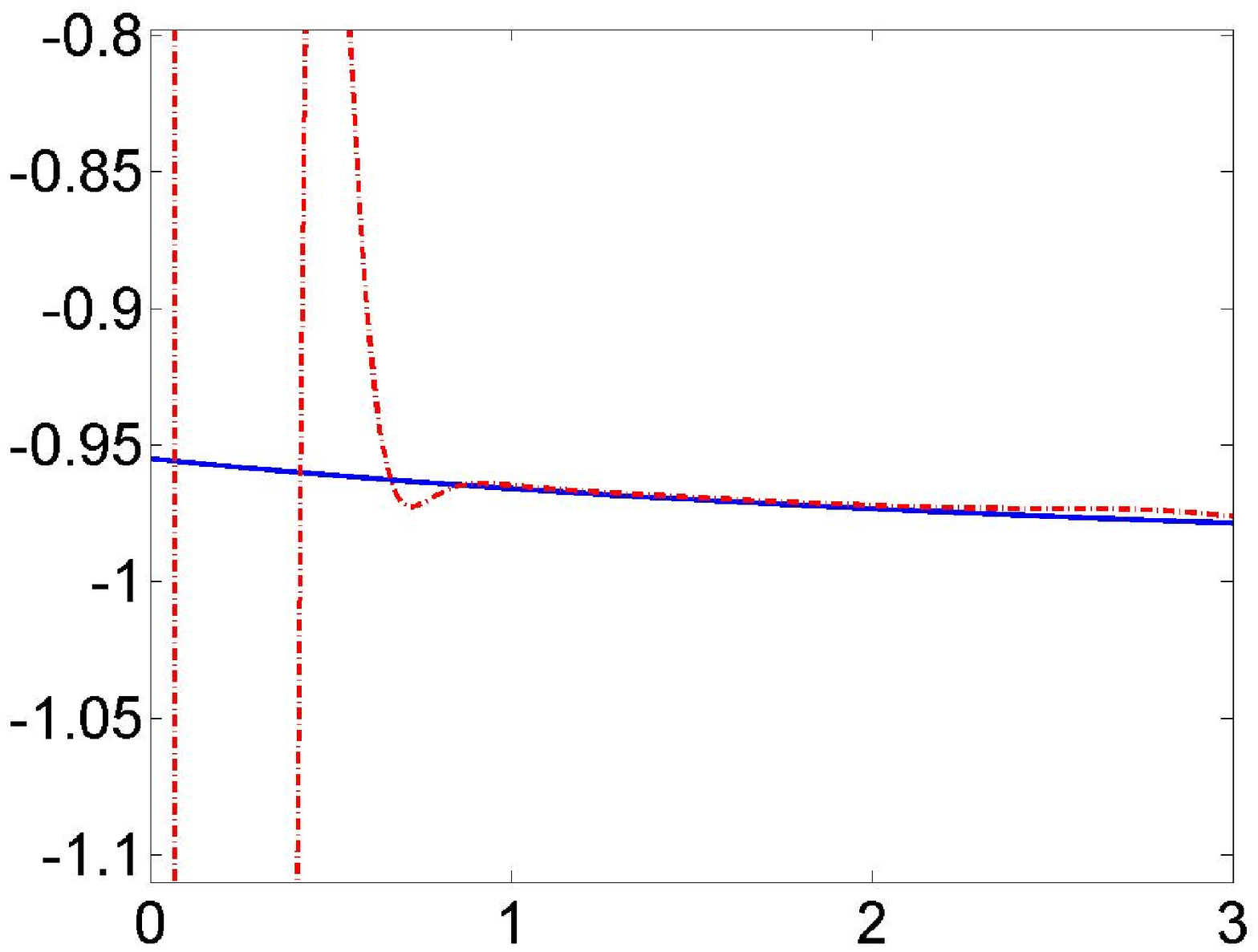}
       \put(-160,10){\rotatebox{90}{parameter $\alpha$ }}
        \put(-37,-10){time}
       \put(-38,90){(b)}
    \end{center}
  \end{minipage}
  \hfill
  \caption{Numerical simulation for $\alpha$ in the underdamping case. The solid line represents the true value of $\alpha$ and the dotted
  line gives the estimate. Plot (b) is a detail of (a) to see better the variation of the estimated
   $\alpha$ in the beginning.}\label{simulaciones alpha}
\end{figure}

\begin{figure}  
  \hfill
  \begin{minipage}[t]{.45\textwidth}
    \begin{center}
      \includegraphics[height=4.4cm]{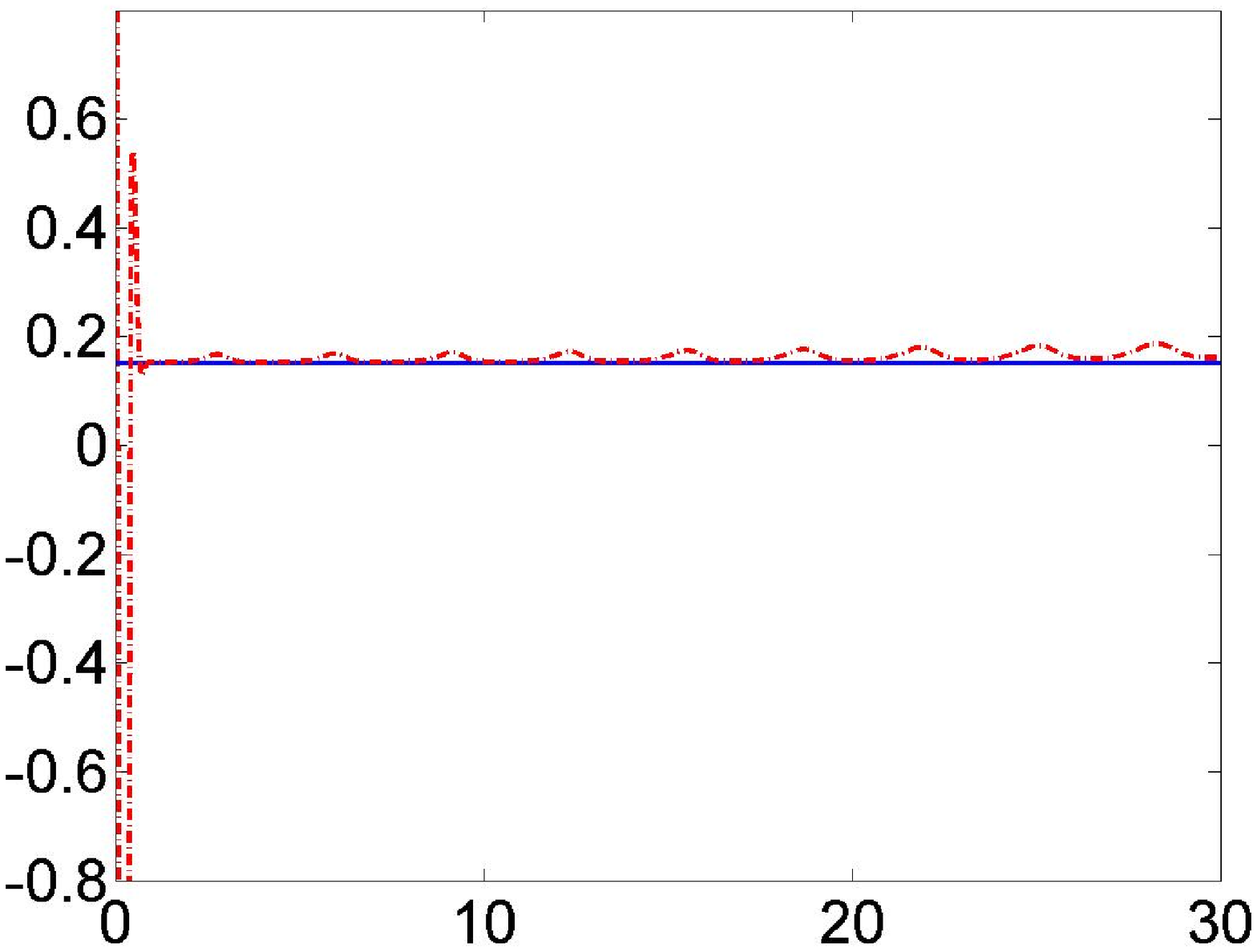}
      \put(-37,-8){time}
       \put(-34,90){(a)}
      \put(-170,10){\rotatebox{90}{parameter $\gamma$ }}
    \end{center}
  \end{minipage}
  \hfill
  \begin{minipage}[t]{.45\textwidth}
    \begin{center}
      \includegraphics[height=4.4cm]{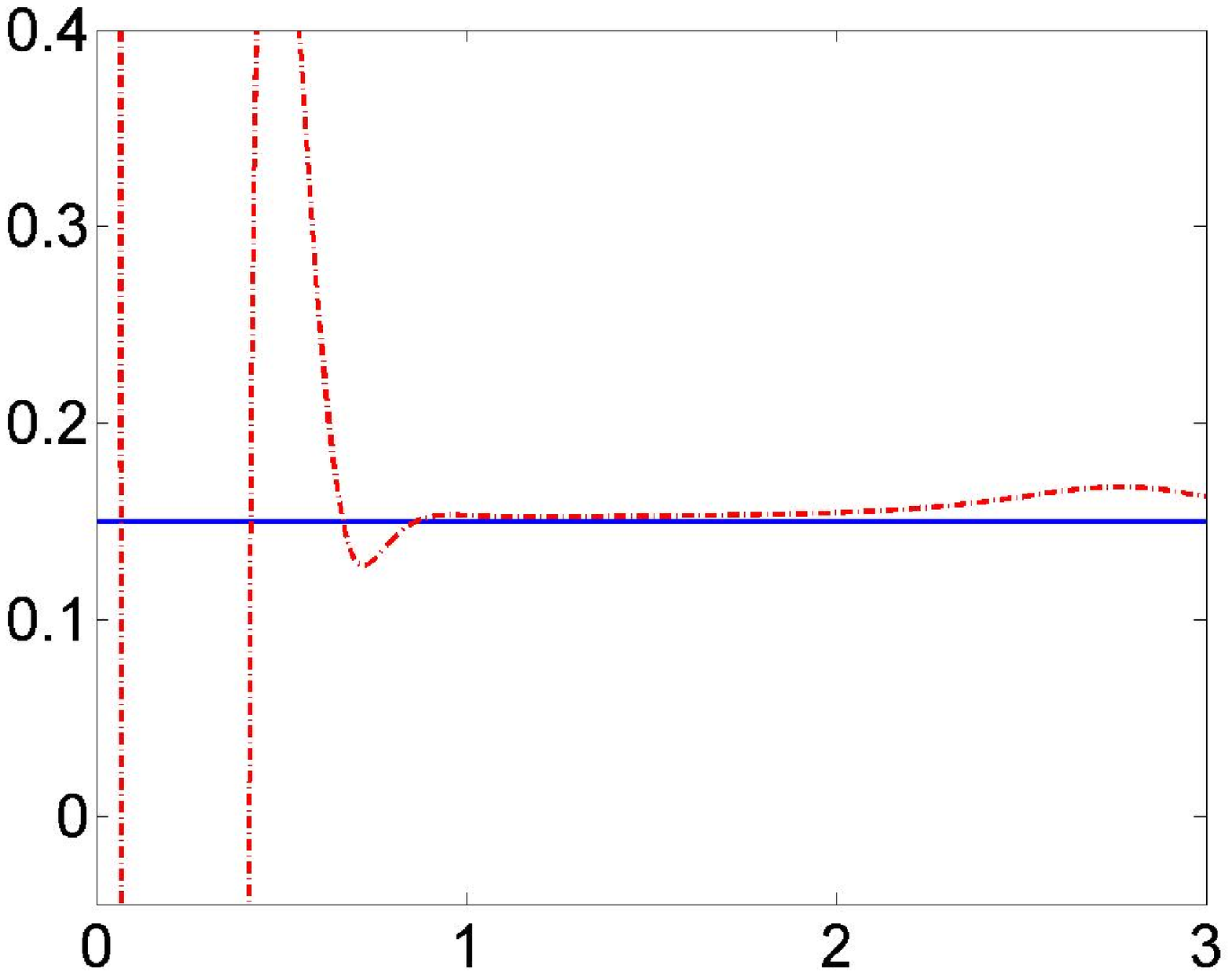}
       \put(-165,10){\rotatebox{90}{parameter $\gamma$ }}
       \put(-37,-8){time}
       \put(-34,90){(b)}
    \end{center}
  \end{minipage}
  \hfill
  \caption{Numerical simulation for the parameter $\gamma$ in the underdamping case. The true value of $\gamma$ is given by the solid line whereas the dotted
  line gives the estimate. 
Plot (b) gives the details of (a) near the origin to better see the variation of the estimate of
   $\gamma$.}\label{simulaciones gamma}
\end{figure}

\begin{figure}  
  \hfill
  \begin{minipage}[t]{.45\textwidth}
    \begin{center}
      \includegraphics[height=4.4cm]{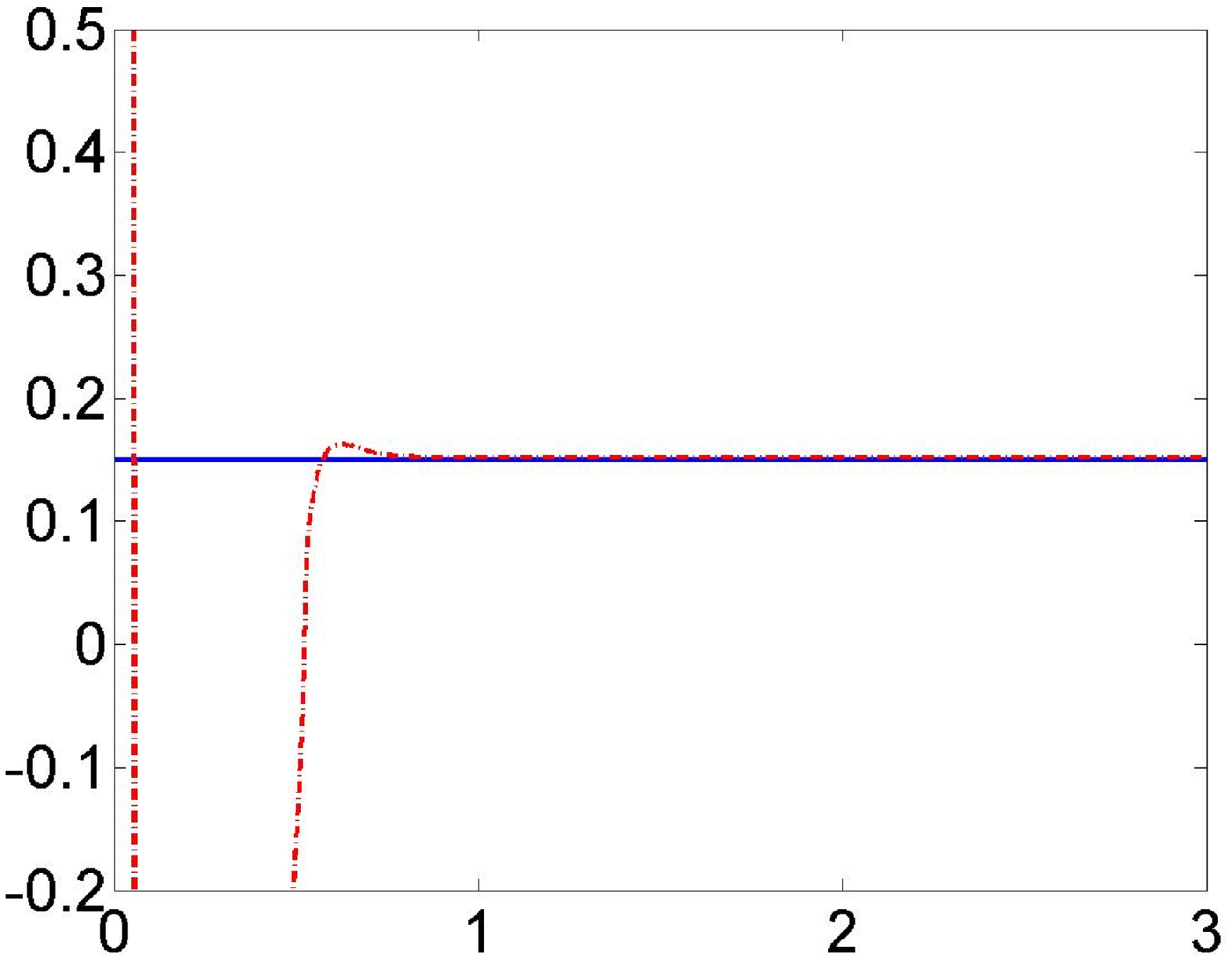}
      \put(-37,-8){time}
       \put(-34,90){(a)}
      \put(-170,10){\rotatebox{90}{parameter $\gamma$ }}
    \end{center}
  \end{minipage}
  \hfill
  \begin{minipage}[t]{.45\textwidth}
    \begin{center}
      \includegraphics[height=4.4cm]{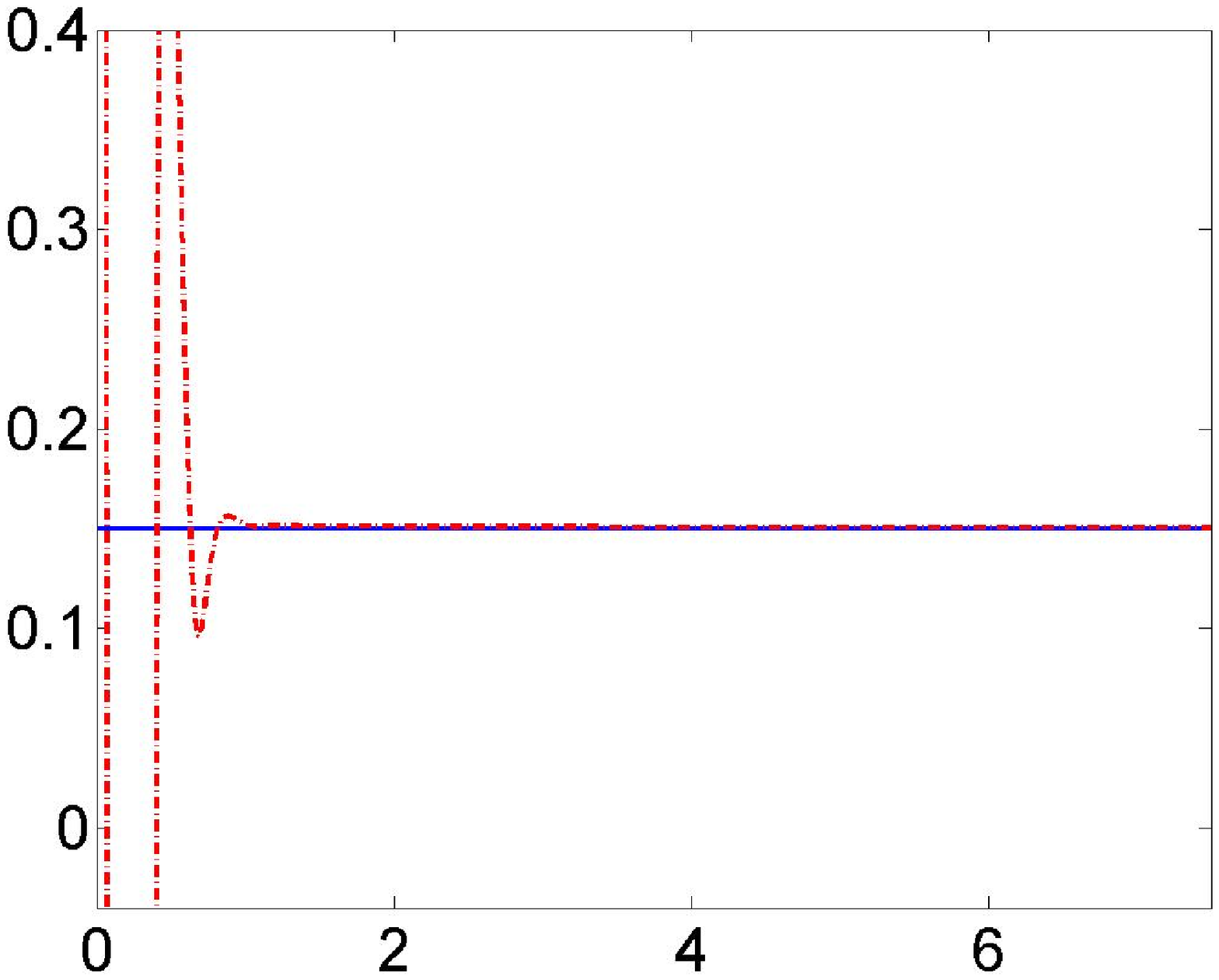}
       \put(-165,10){\rotatebox{90}{parameter $\gamma$ }}
       \put(-37,-8){time}
       \put(-34,90){(b)}
    \end{center}
  \end{minipage}
  \hfill
  \caption{Numerical simulations for the parameter $\gamma$ in the overdamped case [plot (a)], and in the critical case [plot (b)]. The solid line gives the true value of $\gamma$ and the dotted line gives its estimate.
}\label{simulaciones gamma}
\end{figure}

\section{Conclusion}

The adaptive observer scheme of Besan\c con et al is an effective
procedure to rebuild the Riccati parameter of the class of damping
modes of Rosu and Reyes. Despite the fact that in the estimation
algorithm we used an associated time-varying quantity for which the
exponential convergence is not guaranteed, we obtain excellent
results for the estimates of the unknown constant Riccati parameter.
As it can be seen from this work, the application of Besan\c con's
adaptive scheme is not necessarily restricted to the estimation of
constant parameters but one can go with it to some time-varying
cases.  In addition, the scheme provides very good estimates for the
unknown states of this class of oscillators for each of the three
possible cases. Moreover, the rapid identification of the additional
parameter $\gamma$ of the chirped damped oscillators by this method
opens a way to the experimental study of these oscillators.


%

\end{document}